# Investigation of flow field of clap & fling motion using immersed boundary coupled lattice Boltzmann method

Pradeep Kumar S, Ashoke De[*], Debopam Das

Department of Aerospace Engineering, Indian Institute of Technology, Kanpur, 208016, India

*Corresponding author, E-mail: ashoke@iitk.ac.in, Ph: +915122597863, Fax : +915122597561

**Abstract:** This paper deals with the investigation of flow field due to clap and fling mechanism using immersed boundary coupled with lattice Boltzmann method. The lattice Boltzmann method (LBM), an alternative to Navier Stokes solver, is used because of its simplicity and computational efficiency in solving complex moving boundary problems. Benchmark problems are simulated to validate the code, which is then used for simulating flow over two elliptic wing of aspect ratio 5 performing clap and fling flapping motion for different flow parameters such as Reynolds number (Re = 75,100,150), advance ratio (J = 10E-3,10E-2,2) and frequency (f = 0.05Hz,0.25Hz). Numerical simulation is able to capture typical low Reynolds number unsteady phenomena such as, 'wake vortex wing interaction', 'Kramer effect' and 'delayed stall'. The results are both qualitatively and quantitatively consistent with experimental observation. The parametric study involving different combinations of Re, f and J depict distinctly different aerodynamic performances providing physical insights into the flow physics. It is observed that a combination of low f, low J and high Re flow results in better aerodynamic performance. Pronounced lift enhancement via leading edge vortices are obtained in unsteady regime (J<1) compared to quasi-steady regime (J>1). The role of leading edge vortices in enhancing lift are investigated by studying the size and strength of these vortices for different flow conditions. For a given Re, the magnitude of maximum lift coefficient decreases with increasing f irrespective of the value of J; while the same is enhanced with the increasing Re.

## Nomenclature:

| | |
|---|---|
| LBM | = Lattice Boltzmann Method |
| IBM | = Immersed Boundary Method |
| $f_i$ | = Probability Distribution function at $i^{th}$ node |
| $f_i^{eq}$ | = local Boltzmann-Maxwellian distribution function at $i^{th}$ node |
| $F_i$ | = Forcing term |
| $\tau$ | = Relaxation time |
| $\delta(r)$ | = Dirac Delta function |
| $A(t)$ | = Amplitude |
| $\alpha(t)$ | = angle of attack |
| $\beta$ | = sweep angle |
| $\phi$ | = phase angle |
| c | = Chord (lattice units) |
| f | = Frequency (Hz) |
| J | = Advance Ratio |
| $C_L$ | = Coefficient of lift |
| $C_T$ | = Coefficient of thrust |
| $C_D$ | = Coefficient of Drag |

## 1. Introduction

The enigmatic and elusive nature of insect and bird flight has captivated human minds for thousands of years. The evolutionary history of terrestrial animals shows that wings for powered flight, i.e. flapping motion to produce lift and thrust, are produced only for insects, pterosaurs, birds and bats in 450 million years [1]. Insects came into evolutionary scene about 350 million years ago and established itself as the most agile and maneuverable creatures for their size today in addition to the fact that they are the most abundant species in nature [2]. The MAV designers and aerodynamicists have become a recent audience of insect flights because of its power efficient, agile, maneuvering and hovering capabilities. These are quintessential requirements in mimicking the biological beings.

Weis-Fogh [3] on investigating how insects hover in still air, proposed that the insects use a mechanism called "clap and fling" previously unknown to aerodynamicists to enhance the lift generation. Lighthill [4] proposed an inviscid 2D theory for calculating the circulation around the wing undergoing clap and fling motion. Maxworthy [5] pointed out that the magnitude of circulation during fling is much higher than predicted by Lighthill due to the leading edge separation bubble. Spedding, et al. [6] analyzed the effects of altering the wing opening time history and initial angle of wing pair on circulation and lift generation. Lehmann et al. [7] used mechanical model of the fruit fly, to investigate force enhancement. The clap and fling augmented total lift production by up to 17%.

With the advent of powerful numerical techniques, CFD simulations are providing new insights into this complex problem, despite the fact that these simulations heavily rely on empirical data and relevant kinematic input. Since past decade, the LBM has emerged as powerful numerical technique for simulating fluid flows and modeling the physics of fluids. The LBM is based on the kinetic theory which provides the equation for the evolution of discrete molecular velocity distribution functions on uniform Cartesian lattices with diagonal links. The hydrodynamic variables are recovered from computing the moments of the discrete distribution functions.

The immersed boundary method (IBM) coupled with this LBM method provides a very good platform for simulating complex moving boundary problems. In the IBM, the flow field is represented by a set of Eulerian points (Cartesian points) and the boundary of immersed object by a set of Langragian points. The IBM treats the physical boundary as deformable with high stiffness. The restoring force, which is yielded due to the small distortion of immersed boundary, is then distributed into the Eulerian points [8]. These forces are represented through a forcing term by modifying LBM equation like the one suggested by Guo et al. [9]. These forcing terms themselves are calculated from force densities whose expressions are determined by the kind of flow physics that needs to be resolved. For example, in the work of Feng et al. [10], forces like gravity and inter particle forces dominate, while elastic forces dominate as reported by in Hao et al. [11]; and hence the mathematical formulation remains unique in nature. Similarly the immersed boundary forcing can be done both explicitly (direct) and implicitly. Aono et al. [12] used LBM to simulate the vertically oscillating flat plate, while Gao et al. [13] studied the insect hovering flight in ground effect using direct forcing immersed boundary method. Recently, Zhang et al. [14] worked on a locomotion of a passively flat plate with certain flexibility and Ota et al. [15] modeled a two dimensional symmetric flapping wing using direct forcing immersed boundary method.

Very few numerical studies have reported on the low Reynolds number unsteady phenomena such as 'wake capture', 'Kramer effect' and 'delayed stall' in details relating their role in clap and fling motion, which are important physical mechanisms for successful flight of insects and hence provides useful design guidelines for MAVs. Out of which, a handful of them reported the parametric studies on wing kinematics. Notably, Liu et al. [16] have studied the variations of stroke amplitude (from 2 to 5) and angle of attack ($30°$ to $60°$) in Re range of 25 to 100 using multi-block LBM. Miller et al. [17] performed parametric studies using Navier Stokes coupled immersed boundary method where they studied clap and fling motion of single and double wing over a Re range of 8 to 128. However, none of these studies reported the detailed flow physics on these low Re phenomena using IBM coupled with LBM methodology. A deeper understanding of the kinematics, time evolution of vortical structures and flow parameters (Re, frequency, and advance ratio) are quintessential to develop the knowledge base for MAVs. Thus the primary objective of the present study is to throw light on the role of Re, frequency and advance ratio in determining the aerodynamic performance of the clap and fling motion of insect flights using implicit IB-LBM model. The parameters studied herein are as follows: Reynolds number Re from 75 to 150, frequency 0.05 Hz and 0.25 Hz, Advance ratio J from $O(10E-3)$ to $O(1)$.

## 2. Numerical Methods

In this section, we present the details of numerical methods that have been used in the present work.

### 2.1 Lattice Boltzmann Method

The lattice Boltzmann equation (LBE) is the discretized version of continuous Boltzmann transport equation. The solution domain needs to be divided into lattices. D2Q9 (2-Dimensional, 9 speed Quadrature) arrangement is considered to distribute and stream the distribution functions. This equation is assumed to be valid along specific directions & linkages to simulate the fluid flow problem by tracking the evolution of particle distribution function. The corresponding equations of LBM are,

$$f_i\left(x+c_i\Delta t, t+\Delta t\right) = f_i\left(x,t\right) - \frac{1}{\tau}\left[f_i\left(x,t\right) - f_i^{eq}\left(x,t\right)\right] \qquad (1)$$

The hydrodynamic variables are obtained from the moments of distribution function

$$\rho = \sum_i f_i \;,\; \rho u = \sum_i c_i f \quad where \;\; i = 1, 2, ..9 \qquad (2)$$

The lattice sound speed $c_s = \frac{c_i}{\sqrt{3}}$ and correlation between relaxation time $\tau$ and viscosity $v$ is given by

$$\tau = \frac{3v\Delta t}{\Delta x^2} + 0.5 \qquad (3)$$

The local Maxwell equilibrium distribution function, expanding through Taylor series, can be written as

$$f_i^{eq} = \rho w_i \left(1 + \frac{c_i \cdot u}{c_s^2} + \frac{1}{2}\frac{(c_i - u)^2}{c_s^2} - \frac{1}{2}\frac{u}{c_s^2}\right) \quad (4)$$

where, the weighing functions are given by

$$\begin{cases} w_0 = 4/9 \\ w_1 = w_2 = w_3 = w_4 = 1/9 \\ w_5 = w_6 = w_7 = w_8 = 1/36 \end{cases} \quad (5)$$

Unlike Navier Stokes equation, the pressure can be directly found through the following expression without resorting to complex iterative methods.

$$p = \rho/3 \quad (6)$$

The boundary conditions are set such that flow enters the domain with uniform velocity and leaves it at atmospheric pressure. The walls are defined through traditional bounce back condition.

**2.2 Coupling IB with LBM**

An implicit velocity correction Immersed Boundary (IB) condition is implemented in order to satisfy the no slip condition at the moving boundary points. The formulation is adopted from Wu et al. [18]. The same is described in detail below.

We start with the lattice Boltzmann equation (Eq. (1)) which includes a forcing term as proposed by Guo et al., [9]

$$f_i\left(x+c_i\Delta t,t+\Delta t\right)=f_i\left(x,t\right)-\frac{1}{\tau}\left[f_i\left(x,t\right)-f_i^{eq}\left(x,t\right)\right]+F_i\delta t \tag{7}$$

where the forcing term is related as:

$$F_i=\left(1-\frac{1}{2\tau}\right)w_i\left(\frac{c_i-u}{c_s^2}+\frac{c_i.u}{c_s^4}\right).F \tag{8}$$

The hydrodynamic equations are modified to include a force density term.

$$\begin{cases}\rho=\sum_i f_i \\ \rho u=\sum_i c_i f_i+\frac{1}{2}F\,\delta t\end{cases} \tag{9}$$

where F is force density. The Eq.(9) suggests that the total velocity can be seen as sum of two velocity terms, One is contributed from the density distribution function and another is contributed from the force density.

$$u(x,t)=u^*(x,t)+\delta u(x,t) \tag{10}$$

where in $u^*(x,t)$ is intermediate velocity & $\delta u(x,t)$ is velocity correction as defined:

$$\begin{cases}u^*=\frac{1}{\rho}\sum_i c_i f_i \\ \delta u=\frac{\Delta t}{2\rho}F\end{cases} \tag{11}$$

The force density F is determined in such a way that the velocity at the boundary point interpolated from u satisfies the non-slip boundary condition as reported by Wu et al. [18].

In the IBM, the boundary of rigid body is represented by a set of Lagrangian points $X_B^l(s,t), l = 1, 2, ......., m$ as shown in Fig. 1. Here, we can set an unknown velocity correction vector $\delta u_B^l$ at every Lagrangian point. The velocity correction $\delta u(x,t)$ at the Eulerian point can be obtained by the following Dirac delta function interpolation.

$$\delta u(x,t) = \int_\Gamma \delta u(X_B,t)\delta(x - X_B(s,t))ds \tag{12}$$

where

$$\delta(x - X_B(s,t)) = D_{i,j}(x_{i,j} - X_B^l) = \delta(x_{i,j} - X_B^l)\delta(y_{i,j} - Y_B^l) \tag{13}$$

$$\delta(r) = \begin{cases} \dfrac{1}{4}\left(1 + \cos\left(\dfrac{\pi |r|}{2}\right)\right) & |r| \leq 2 \\ 0 & |r| > 2 \end{cases} \tag{14}$$

The Eq. (12) can be approximated as

$$\delta u(x_{i,j},t) = \sum_l \delta u_B^l(X_B^l,t) D_{i,j}(x_{i,j} - X_B^l)\Delta s_l \tag{15}$$

where $\Delta s_l$ is the arc length of the boundary element. The fluid velocities at Eulerian points can be corrected using following equation,

$$u(x_{ij},t) = u^*(x_{ij},t) + \delta u(x_{ij},t) \tag{16}$$

To satisfy the non-slip boundary condition, the fluid velocities at the boundary points obtained by interpolation using the smooth delta function must be equal to the wall velocities $U_B^l(X_B^l,t)$ at the same position

$$U_B^l(X_B^l,t) = \sum_{i,j} u(x_{i,j},t) D_{i,j}(x_{i,j} - X_B^l) \Delta x \Delta y \qquad (17)$$

This can be rewritten as

$$U_B^l(X_B^l,t) = \sum_{i,j} u^*(x_{i,j},t) D_{i,j}(x_{i,j} - X_B^l) \Delta x \Delta y + \\ \sum_{i,j}\sum_{l} \delta u_B^l(X_B^l,t) D_{i,j}(x_{i,j} - X_B^l) \Delta s_l D_{i,j}(x_{i,j} - X_B^l) \Delta x \Delta y \qquad (18)$$

The above-mentioned equation can be further re-written as the following matrix form, AX=B

where,
$$\begin{cases} X = \{\delta u_l^1 \quad \delta u_l^2 \quad , \quad . \quad . \quad , \quad \delta u_B^m\} \\ B = \{\Delta u_1 \quad \Delta u_2 \quad , \quad . \quad . \quad , \quad \Delta u_m\} \\ \Delta u_l = U_B^l(X_B^l,t) - \sum_{i,j} u^*(x_{i,j},t) D_{i,j}(x_{i,j} - X_B^l)\Delta x \Delta y \end{cases} \qquad (19)$$

By solving the above equation at every time step, one can obtain the unknown velocity correction at all the Lagrangian boundary points (Fig.1). Through Dirac delta interpolation as explained above, one can obtain the Eulerian velocities and hence force density. This is then used to obtain the new equilibrium distribution at each time step.

## 3. Results and Discussion

In this section detailed flow field during clap and fling motion for different flow parameters are presented after validating the code. The LBM code used here has been extensively validated in previous work [19] and [20]. Furthermore, in order to demonstrate the validity of immersed boundary coupled lattice Boltzmann method, benchmark problems like flow over stationary cylinder and wing performing a figure eight flapping motion are simulated and reported before moving to the detailed simulations.

**3.1 Flow over a rigid stationary circular cylinder:**

A lattice domain of 50D x 50D and a blockage ratio of 50 are chosen for the simulation as reported in the literature. The boundary conditions are set such that the flow enters the domain at uniform velocity and leaves the same at atmospheric pressure using Zou et al. [21] formulation. As in the earlier the case the no slip conditions on the walls are applied through the bounce back scheme. The no slip condition to the cylinder boundary is given through the Eq. (17) which calculates the velocity corrections needed to apply the no slip condition which is then converted into force density, the unknown quantity in the forcing term of Boltzmann equation. The vortex shedding frequency is calculated from the FFT of instantaneous velocity obtained from the simulation. The drag coefficient predictions are in closer range with the experimental data and other numerical results (Fig.2A). The Strouhal numbers calculated from the shedding frequency are plotted against Reynolds number in Fig.2B.

**3.2 Moving boundary problem using IBM+LBM**

The unsteady flow arising due to the flapping motion is compared with Wang's [22] work by simulating the flow with similar parameters and kinematic patterns as reported by them.

An elliptic wing of aspect ratio (chord to thickness ratio) 7.5 is made to undergo translational and rotational motion. The wing pitches about its center while sweeping in stroke plane which is inclined at an angle β. The motion of the wing center is governed by the following kinematic motion,

$$A(t) = \frac{A_0}{2}\left(\cos\left(\frac{2\pi t}{T} + \phi\right)\right) \qquad (20)$$

$$\alpha(t) = \frac{\pi}{4} - \frac{\pi}{4}\left(\sin\left(\frac{2\pi t}{T} + \phi\right)\right) \qquad (21)$$

where, $\phi$ is phase difference. The following parameters are chosen from Wang [22].

$$\frac{A_0}{c} = 2.5, \ f = 40 Hz, \ \beta = \pi/3, \ \phi = 0$$

Here A(t) is time dependent amplitude, $\alpha(t)$ is angle of attack, T is time period, $A_0$ is peak amplitude and f is frequency.

The Reynolds number is defined based on the maximum tip velocity, where the tip velocity is function of amplitude and chord of the elliptic wing as:

$$\text{Re} = \frac{U_{tip} \cdot c}{\nu} = \frac{\pi f A_0 c}{\nu} \qquad (22)$$

where, $U_{tip}$ is wing tip velocity and $\nu$ is kinematic viscosity. A lattice domain of 50c x 25c is chosen for this simulation. The present simulations capture the vortex dynamics as reported by Wang [22]. During the translational phase of the wing motion, a pair of leading edge vortices (LEV's) & trailing edge vortices (TEV's) is created in opposite direction (Fig.3A). When the rotational phase starts, vortices combine to form a dipole (Figs.3B-3C); which is ejected together as vortex pairs at the end of the cycle (Fig.3D). The time histories of force coefficients attains periodicity within first few cycles and is found to be matching the trends as reported by the Wang [22]. The increase in lift coefficient corresponds to formation of leading edge vortex (Fig.3E). The small phase shift in Fig.3E between the present steady and Wang's [22] work is attributed to the fact that initial positional angle in present study is different from Wang's work. As the initial positional angle was not explicitly mentioned, we assumed it and carried out the simulation. Despite this small phase shift, the IBM coupled LBM simulation could capture the exact flow physics of figure of eight motion as mentioned by Wang [22].

**3.3 Clap & Fling motion:**

The present study is to understand the flow physics behind the clap and fling motion in insect flights and study the effects of flow parameters such as advance ratio (J) and frequency (f) on the dynamics of this motion at different Reynolds numbers. Before continuing further, it is noteworthy to mention that the clap & fling, as the name suggests involves two different phases namely 'clap' and 'fling'. Fig.4 depicts the three important stages in the motion both in 3D (Figs.4A – 4C) and 2D (Fig.4D – 4F) where the freestream direction is depicted by streamlines in Fig.4D. Initially the leading edge of both wings claps together either partially or totally at the end of the upstroke (Fig.4A). During the 'fling' phase the wings rotate around their trailing edges, thus flinging apart. This second phase appears before the commencement of the down stroke i.e. the clap & fling appears at the stroke reversal. Fig.4B presents the start of that fling cycle which at the end of its stroke reverses (Fig.4C) and comes back to clap position. In between the start & end of the fling there is a translation motion where the wing flaps its wing out of plane in a real 3-dimensional motion.

The main difference between the three dimensional and two dimensional clap and fling motion is that of the spanwise flow during the motion which stabilizes the leading edge vortices delaying the stall and hence preventing build of drag due to flow separation at higher angles of attack. Interestingly, a large number of literature shows this flow phenomenon is not prominently reflected in the aerodynamic forces measured, i.e. the forces measured in 2D as well as 3D matches quite well [23]. Thus based on this observation, one can safely justify neglecting the out of plane translation motion for 2D simulation.

To study the sensitivity of the simulated flow physics to the grid resolution and domain size, Re = 75 is used. Reynolds number is defined using Eq. (22).Three grid dimensions are chosen. Domain 1 has a dimension of 50c*25c while domain 2 and 3 have dimensions 50c*50c and 60c*40c respectively. Fig.5A shows the time history of lift for these three domains. The differences between the computed lift coefficients are found to be small. In fact there is almost no difference between the lift coefficients computed using domain 1 and domain 2. Figs.5C-5E shows that the streamline plot of wing performing clap and fling are qualitatively same in different domains. Thus to save the computational time domain 1 (50c*25c) has been selected for the present simulation.

Similarly Fig.5B shows the time history plot of lift for finer and coarser grids. Simulations are run for same Re (=75) with grids refined by a factor of 2. One can see that in fine grid case, there is very slight increase in lift co-efficient compared to the coarse grid while the qualitative pattern of the cycle remaining the same. Figs.5F-5G is a qualitative description of wing performing clap and fling using streamline plot. Since both the coarse and finer grid are comparable both in qualitative and quantitative terms, we choose the coarser grid to save the computational time.

In order to replicate the clap and fling motion, the following kinematic motion is given to two rigid elliptic foils,

$$\alpha(t) = \frac{\pi}{3}\left(\sin\left(\frac{2\pi t}{T} + \phi\right)\right) \qquad (23)$$

where ϕ is phase difference which is set to zero, $\alpha(t)$ is the angle of attack which changes from $0°$ to $60°$ and T is the time period taken to complete one clap and fling cycle. The flow parameter advance ratio (J) is defined as ratio between free stream velocity ($U_\infty$) and wing tip velocity ($2\Phi fc$). Here $\Phi$ is maximum opening angle which is $120°$.

$$J = \frac{U_\infty}{2\Phi fc} \qquad (24)$$

The boundary conditions are set such that flow enters the domain with uniform velocity and leaves at atmospheric pressure. The elliptic foils are hinged at the trailing edge so that the rotational motion takes place about the trailing edge. The gap between the two foils c/2 is chosen based on the constraint that the Eulerian nodes surrounding the Lagrangian nodes of each wing does not coincide or interfere with other.

The simulation is performed for three different Reynolds numbers (75,100,150) with three different advance ratios J (1.67*10E-3, 1.67*10E-2, 2) and for two different frequencies f (0.05Hz, 0.25Hz). The rotational velocities of the moving boundaries are given in addition to the kinematic motion. This boundary velocity ensures the acceleration and deceleration of the elliptic foils as it undergoes the clap and fling motion.

While Re and J corresponds to the physiology ranges of real insects, frequency is chosen based on the limitations of the numerical model; e.g. Re of fruitfly is 120-130 that is within the range of Re considered in the simulations. The advance ratio (J) corresponds to different flow regimes namely steady and unsteady. Similarly the flapping frequency of fruitfly is around 250 Hz. Owing to numerical stability issues, these real frequency ranges cannot be handled. Hence based on those limitations, frequencies are chosen as 0.05 Hz and 0.25 Hz.

The streamlines of the flow around the wings performing clap and fling motion are shown in the Fig.6 for selected time instances corresponds to the case of Re=100, f=0.05Hz, J=1.67E-3. The Figs.6A–6I present the streamline plots of half stroke of a clap and fling motion. Figs.6J–6K show corresponding time histories of lift and drag coefficients. Similarly Figs.6L–6M represent the leading edge and trailing edge circulation histories. Initially the wings are in clap position (Fig. 6A).

At the beginning of the down-stroke, the wings fling apart from clap position about their trailing edge. This generates two attached LEVs which create regions of low pressure above the wing and thus generating lift. These LEVs are fed by rotational motion, making it grow larger with time until it sheds away from the wing. Thereafter, the LEVs are shed at the end of the stroke reversal (Figs.6B–6E). While the wings perform clap motion, the vortices that shed during the fling motion interact with the wing. These shed vortices being high energy structure with strong vortex field aids in further enhancing the lift as the wing encounters it (Figs.6F–6G). This phenomenon is referred as wake capture or wake wing interaction. This happens because of the rapid wing movement which encounters the shed vortices before they move along the wake. This is a remarkable phenomenon because of the fact that insects use this mechanism to extract the lost energy from its own wake thereby increasing the overall efficiency of the lift force production.

At the end of the clap motion the leading & trailing edge vortices are shed (Figs.6H–6I). No TEVs are shed during the fling phase. The LEVs are stronger than the TEVs throughout the stroke. This also attributes to the higher lift forces encountered in clap and fling motion.

The time histories of lift and drag coefficients of corresponding case is illustrated in Figs.6J–6K. The large lift peak observed in the Fig.6J is due to the formation of LEV. As LEV grows in strength with time and remained attached in this unsteady motion that results in the delay of stall, enhanced lift is achieved. The small peak that immediately follows the large peak is due to the wake capture. This is can be verified from Fig.6G showing two vortices interacting with the wing surface. One can notice peak in drag coefficient in Fig.6K at the end of down-stroke which is primarily due the ejection of TEVs. The ejection of LEVs and TEVs can also be seen from the time histories of circulation plotted in Figs.6L–6M.

**3.4 Effects of Advance Ratio (J) and Frequency (f) at constant Re:**

The size and strength of the LEVs is a function of advance ratio. For lower frequency and High advance ratio (J>1), the LEVs are poorly formed compared to the low frequency and low advance ratio (J<1) case. But for high frequency and high advance ratio (J>1), the elliptic foils generates LEVs of size comparatively lesser than that of flow parameters involving lower advance ratio (J<1). Fig.7 presents the case of clap and fling mechanism involving different advance ratio and frequency at Re=75.

It is to be noted that all the subplots in Fig. 7 corresponds to the same time instant (t/T =0.55) of the stroke cycle. Fig. 7(a) corresponds to low frequency f =0.05Hz and advance ratio J=2. One can notice that leading edge vortices are poorly formed. The non-dimensional size (L) of the vortex core is 2.00 and the non-dimensional circulation ($\Gamma$) is 0.5478. This is attributed to the higher incoming flow velocity (compared to flapping velocity) which destabilizes the leading edge vortex formation. It is noteworthy to mention that the circulation here is non-dimensionalised through the following expression:

$$\Gamma = \frac{\Gamma^*_{la}}{U_{tip}c} \qquad (25)$$

where "la" denotes lattice units.

Fig. 7(b) Corresponds to the case of low frequency f = 0.05Hz and low advance ratio J =1.67E-3. The leading edge vortices are clearly formed and larger in both the size (L =9.0553) and strength ($\Gamma$ = 1.639) as compared to Fig. 7(a). Similarly Fig. 7(c) corresponds to high frequency f = 0.25Hz and advance ratio J=2. The leading edge vortices in this case is clearly formed with much larger size (L = 12.5499) and greater strength ($\Gamma$ = 1.001) compared to the Fig. 7(a). Fig. 7(d) which represents high frequency f =0.25Hz and low advance ratio J=1.67E-3 has an LEV of size (L = 11.6190) and strength ($\Gamma$ = 0.9561) comparable to that of Fig.7(c). The trend is same throughout the clap and fling cycle. This is reflected even in the maximum lift coefficient data plotted in Fig.11, section G. In all advance ratios, the wing which flaps with low frequency has higher lift-coefficient compared to the wing that flaps with high frequency. This is due to the fact that if J is constant and frequency is decreased the forward velocity decreases resulting increase in lift co-efficient. The circulation data in Fig. 7(b) and Fig. 7(d) predicts the same. Fig. 7(a) has no leading edge vortices whereas Fig.7(c) has, still former produces higher lift coefficient.

It is important to note that while frequency is changed, the Re is kept constant by changing the kinematic viscosity. The effect of the frequency is reflected through the wing tip velocity whose relation is given as $U_{tip} = 2\Phi \text{f} c$. From Eq. (22), Re is directly proportional to the frequency and inversely to the kinematic viscosity. The basic assumption here in LBM is that physical Reynolds number must be equal to the Lattice Boltzmann Reynolds number. The velocity, kinematic viscosity and chord entering the term LBM Reynolds number are not in physical units but in lattice units. Hence one has the liberty of choosing the three parameters independently without any constraints as long as the LBM Reynolds number matches physical Reynolds number. Hence to study the effect of different frequency of flapping wing at same physical Reynolds number, the change in frequency is counteracted by changing kinematic viscosity thus keeping the LBM Reynolds number constant.

Thus from above reasoning one can try to explain why in high advance ratio case (J=2), LEV is not formed at lower frequency but appears in higher frequency. Fig 7(c) has higher ν value compared to 7(a) and hence the boundary layer thickness (which is proportional to sqrt(ν) ) of case 2 is higher that results in the formation of LEV.

**3.5 Effect of Re at constant Advance Ratio and Frequency:**

The Fig.8A represents the comparison of size of LEV at different Reynolds number (Re = 75,100,150) and advance ratio (J=2, J=1.67E-3) for constant frequency f=0.25Hz. The figures correspond to same time instant of stroke cycle. Figs.8G–8H represent the vortex size and strength versus Re plot at different advance ratio.

The plot shows that the size of the vortex first increases with Re and then decreases for both the advance ratio. In case of vortex strength, the advance ratio J = 1.67E-3 shows the same trend of vortex size, i.e. first increase and then decrease. But for J = 2, the plot shows a linear decreasing trend, i.e. as the Reynolds number increases the vortex strength reduces linearly. It is interesting to note that the strength of the LEVs are large in advance ratio J=1.67E-3 than J=2.

This reinforces the fact that lift enhancement via leading edge vortex is prominent only in unsteady regime (J<1) not in steady regime (J>1). This is again established through analysis of forces in next section.

**3.6 Comparison to numerical results and experimental data:**

The Fig.9A compares the present clap and fling IB-LBM simulations with the experiment conducted by Gosh et al. [24], involves similar clap and wing motion to analyze force characteristics at Reynolds numbers of order O ($10^3$). The experiments were done using a monarch butterfly shaped model which simulates the flapping motion using a 4 bar quick return mechanisms. They carried out force measurements by placing the model over a two component platform balance inside a wind tunnel. The time history of lift forces for simulation performed at Re=75 is scaled up to Re $\simeq$ 3000 by using the expression $F = \frac{1}{2}\rho C_D S U_{tip}^2$, where S is the surface area, $\rho$ is density, wingtip velocity $U_{tip} = 2\phi R f$ in order to match these experimental data. The red line represents immersed boundary simulation while the black dashed line depicts experimental data. There is an excellent agreement between the experimental and IB-LBM simulation.

Sohn et al. [25] carried out a flow visualization and aerodynamic load calculation of three types of clap and fling motion through experiments and computations. The circulation time history obtained by them in investigating "cyclic fling and clap" motion is compared with the present IB-LBM simulation. Their computational studies were performed by solving 2D Navier Stokes equation for Re = 5600. By scaling up the present study to theirs, a comparison is made in Fig.9B which agrees reasonably well.

Figs.9C–9D compare Wang et al. [23] numerical simulation involving one-winged flapping strokes at Re = 75. The motion of the wing is similar to the motions used in present simulation, except the fact that the stroke used by Wang et al. [23] has a minimum angle of attack of 45° whereas the present study has a minimum angle of attack of 30°. The slight discrepancy in the lift coefficient peak is attributed to these variations. The drag coefficient in Fig.9D though agrees quite well with the simulation.

**3.7 Time Histories of Lift and Thrust Forces:**

The lift and thrust forces are calculated by integrating the force densities over the Lagrangian nodes, i.e. immersed boundary points. These force densities themselves are obtained by rearranging Eq.(11) as shown below:

$$F_x = \frac{2\rho\delta u}{\Delta t}; F_y = \frac{2\rho\delta v}{\Delta t} \qquad (26)$$

Here the terms lift and thrust are used in conventional aerodynamic sense, i.e. the components of resultant force that is perpendicular and parallel to the free stream respectively. Negative thrust is defined as drag. According to the conventions followed, thrust force operates in vertical direction and lift operates in the sideways, i.e. lateral direction.

The time history trends of aerodynamic forces on the elliptic foil performing clap and fling are plotted in Fig.10. The lift and thrust coefficients for a range of advance ratio (J) and frequencies are plotted as function of time for Re =100. The top row corresponds to the cases where the advance ratio is J = 1.67E-3, while the bottom row corresponds to advance ratio J =2. The red and black line represents frequency f = 0.05Hz and f = 0.25Hz respectively. From the Fig.10 one can notice that for a given Reynolds number, the lift coefficients are high for low frequency (f =0.05Hz) than high frequency (f=0.25Hz) provided the fact that advance ratio is same for both the frequencies. As the advance ratio increases to greater than unity (J=2 in the present case), the magnitude of lift coefficient decreases for both high and low frequencies. This is expected because the flow is now in quasi-steady regime where the LEVs don't play any important role in lift generation. For thrust coefficients, at low advance ratio (J=1.67E-3), both the frequencies produces comparable thrusts. At advance ratio $J = 2$, the higher frequency (0.25Hz) produces higher thrust. It is noteworthy to mention that the force histories during down and upstroke look almost identical in the present situation. The reason being is the motion of flapping wing through a cycle is divided into translational and rotational phases, where the translational phases consists of two half strokes namely the down stroke and the up stroke. While the former refers to the motion of the wing from its rearmost position to its foremost position, the latter describes the return cycle. The rotational phase occurs at the end of every half stroke. This is a stroke reversal phase which in this case refers to a specific type of motion called the clap and fling motion (as clearly explained in the section 3.3.) Every half stroke is accompanied by a clap and fling stroke reversal phase, i.e. for each flapping cycle, the phenomena occurs twice, one at the end of the up stroke and other at the end of down stroke. Since this is a 2D motion, the out of plane translational motion cannot be depicted. Hence the upstroke and down stroke of a flapping cycle is filled only with the clap and fling motion. Thus the upstroke and down stroke plot looks identical as expected. The same has been amply justified through the Figures 4, 6 & 9 also.

The maximum and mean lift, thrust coefficients for different flow parameters are plotted against Reynolds number in Figs.11A–11D. The first row corresponds to maximum lift and thrust coefficients against Re. The maximum lift coefficient is clearly a function of Reynolds number. It increases with increasing Re irrespective of advance ratio and frequency. The maximum lift coefficients for advance ratio J=2 at various Reynolds number are very less compared to the low advance ratio regimes (J<1); i.e. with the increasing J value, the maximum lift coefficient is decreasing. This again portrays why insects and MAVs of our time are more aerodynamically efficient in unsteady regime than in steady regime. With increasing frequency the maximum lift coefficient decreases irrespective of advance ratio at a given Re. The maximum thrust coefficients also decrease with increasing advance ratio for a given flapping frequency. They also increase with increasing Re irrespective of advance ratio or frequency. Also with increasing frequency they increase.

The second row illustrates the mean lift and thrust coefficients plotted against Re. The trends are different from those of maximum lift and thrust coefficients. One can notice from Fig.11C that the mean lift coefficient increases with increasing advance ratio. For low advance ratio (J<1), the mean lift coefficient is relatively constant till Re = 100 and then increase at Re =150. In case of high advance ratio, the mean lift coefficient is relatively same with increasing Re; while at low advance ratio, there is an increase in the magnitude. The negative mean thrust coefficients in Fig.11D must be interpreted as mean drag coefficients. At advance ratio less than one, the mean thrust coefficient tends to decrease with increasing frequency. At J =2, the mean drag coefficients remains almost same irrespective of increase in frequency. The mean lift coefficient increases for J = 2, J = 1.67E-3, while it first increases and then decreases at J = 1.67E-2.

Some of the above mentioned trends can be more clearly visualized using Figs.11E–11H which depict the maximum lift & thrust coefficients and the mean lift & thrust coefficients plotted against the advance ratio for different Re and frequency. The more interesting is the mean thrust coefficient plot against the advance ratio (Fig.11H). For advance ratio less than one, for all Reynolds number, the lower frequency produces mean thrust whereas the higher frequency produces mean drag. As the frequency increases there is a mean thrust reversal. The frequency dependency of the thrust coefficients is observed due to the fact that it is determined by the strength of the ejected jet and interacting vortices, hence tip velocity of the wing. Some of these trends are in excellent match with the experiments done by Banerjee et al. [26] and Gosh et al. [24].

## 4. Conclusions

Through this study we have investigated clap and fling motion using immersed boundary coupled lattice Boltzmann method. The computed results demonstrates that the unsteady mechanism is better model to investigate insect flight because of its ability to capture various low Reynolds number phenomena observed in real world. In the present study of clap and fling, the high lift produced during the motion is due to the generation of leading edge vortices. The wake capture amounts for very small percentage in total lift. Similarly the thrust is also generated when the trailing edge vortices are shed during stroke reversal.

An extensive parametric study is conducted to analyze the effects of flow parameter on the aerodynamic performance of the wing motion. The effects of advance ratio and frequency over a range of Reynolds number show that the aerodynamic performance of the clap and fling is a strong function of these variables. This is very well established through time histories of lift and drag coefficients and their variations with above mentioned parameters. The magnitude of lift and thrust in quasi-steady regime ($J>1$) is low compared to the unsteady regime as the leading edge vortices don't play a major role in lift enhancement. The comparison of size and strength of LEV's for flow fields involving different flow parameters provides insights into the force production mechanisms. At a given Re, the magnitude of lift coefficient decreases with increasing frequency. The flapping motion involving higher frequency exhibit wing-wake interaction as expected. Maximum lift is found to be a strong function of Re which on increasing will yield higher lift. The lift which is mainly produced during the down stroke is inversely proportional to the advance ratio. Similarly, the thrust coefficients produced during the clap motion is found to be decreasing with increasing frequency.

Wing flexibility, is an important parameter which is not accounted for in the present study. While the insect wings in reality are flexible, it is treated as rigid body here. This assumption affects our results in two ways. One, under-prediction of lift enhancing capacity of "clap and fling" mechanism. Many existing research works both experimental as well as computational points out that the effect of flexibility is to lower drag and enhance lift [22] [23] [27] [28]. Percin et al. [29] shows that in flexible wing case phenomenon of wake capture plays a prominent role in enhancing lift. Two, the complex interplay between inertial, elastic and aerodynamic forces which is very much essential in understanding the "fluid-structure interaction" and its influence in the resulting flow phenomena is not captured. We hope to cover these issues in our future investigations.

The immersed boundary coupled lattice Boltzmann model used for simulating clap and fling motion is able to capture complex vortex interaction well throughout the cycle with reasonable accuracy. Complete agreement is obtained with the established flow physics both from computational and experimental data available. With a ready provision to include flexibility of the body, this method appears highly promising. The parametric study performed helps in developing a deeper understanding in time evolution of vortical structures which are quintessential for MAV designs.

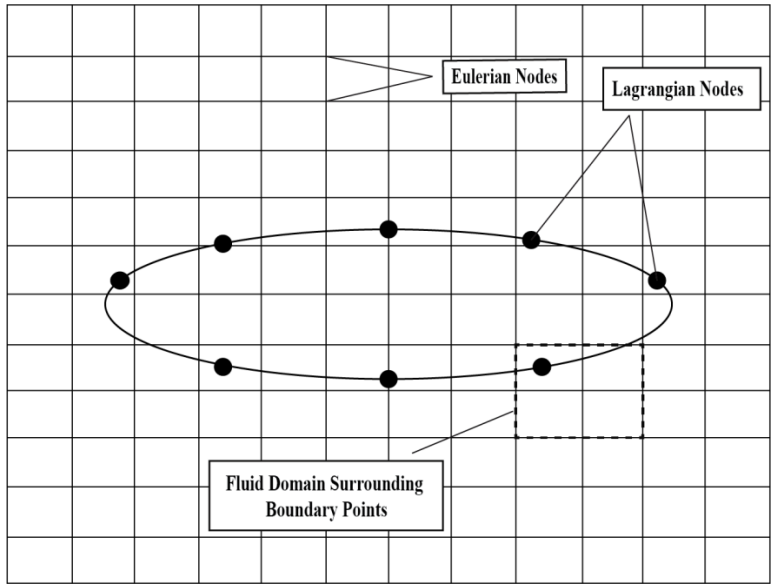

**Fig. 1:** Conguration of Eulerian and Lagrangian Nodes. The domain marked by the dashed line indicates the fluid domain surrounding the Lagrangian boundary point.

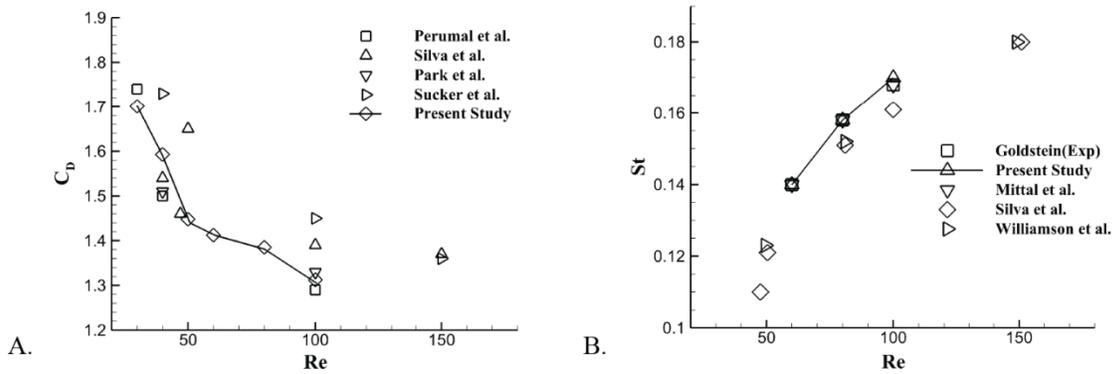

**Fig.2:** Variations of (A) mean drag coefficient, $C_D$ and (B) Strouhal Number, St with Reynolds number for flow past a circular cylinder, showing the validity of the code used.

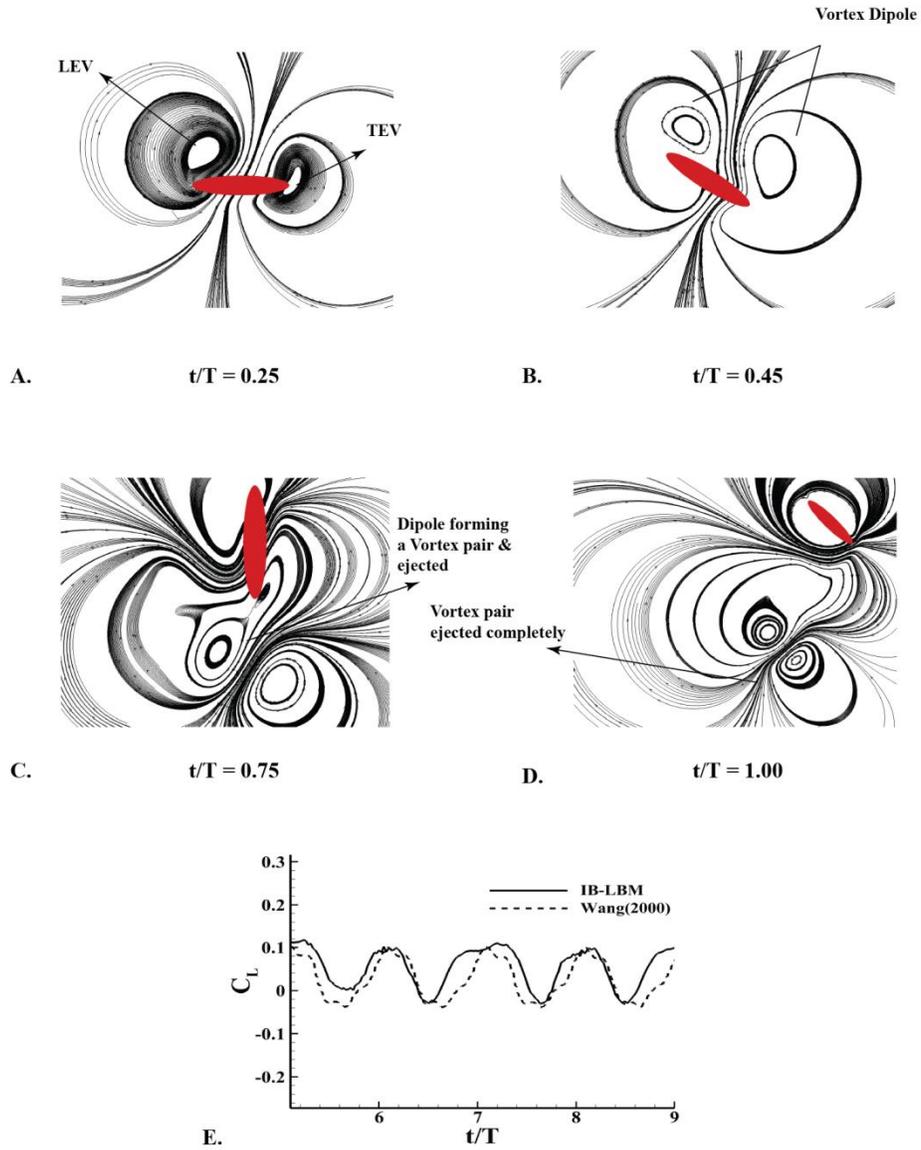

**Fig. 3:** Streamline plot of a Figure of eight motion. A) The wing is in translational phase. Formation of Leading and Trailing edge vortices is shown. B) The rotational phase starts resulting in the dipole formation C) The dipole slowly forms a vortex pair D) The vortex pair is ejected at the end of cycle. E) The increase in lift coefficient at the start of the cycle is due to the formation of Leading edge vortex

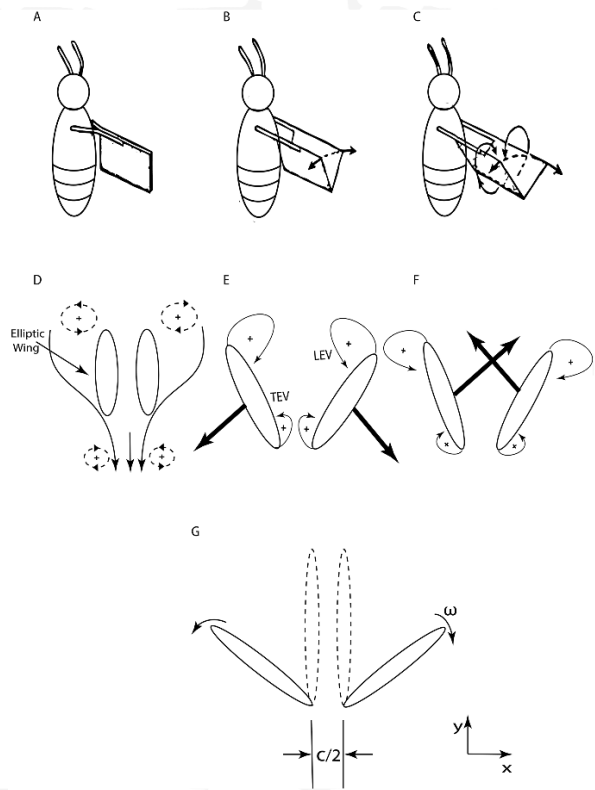

**Fig. 4:** The diagrams (A-C), shows the clap and fling motion in 3D. (D-F) shows the corresponding 2-D approximations of the same phenomenon. The wing initially at clap position (A&D) starts flinging apart (B&E) resulting in the leading and trailing edge vortices. As the wing reversal takes place (C&F), the fling ends with coming to clap position again. The diagram G, is a sketch of two elliptical foil performing clap and fling motion. $\omega$ is angular velocity of the wing of chord c. The gap between the two wings of length c/2. The freestream direction is depicted by streamlines in Fig.D

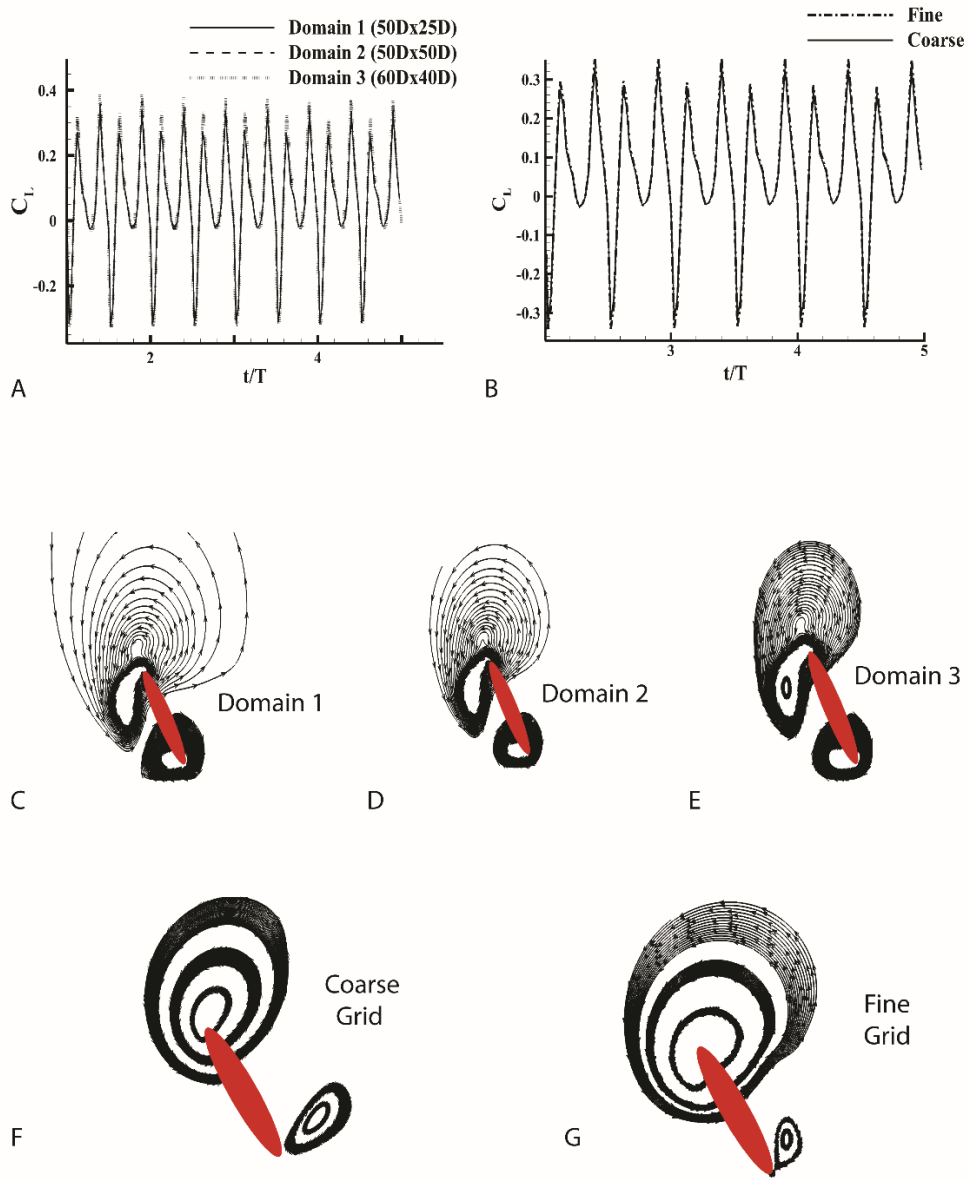

**Fig 5 :** (A) Time history of lift Co-efficient for three different domains. (B) Time history of list Coefficients for fine and coarse grid. (C-E) Qualitative comparison of wing performing clap and fling motion for different domain size. (F-G) Qualitative comparison between coarse and fine grid.

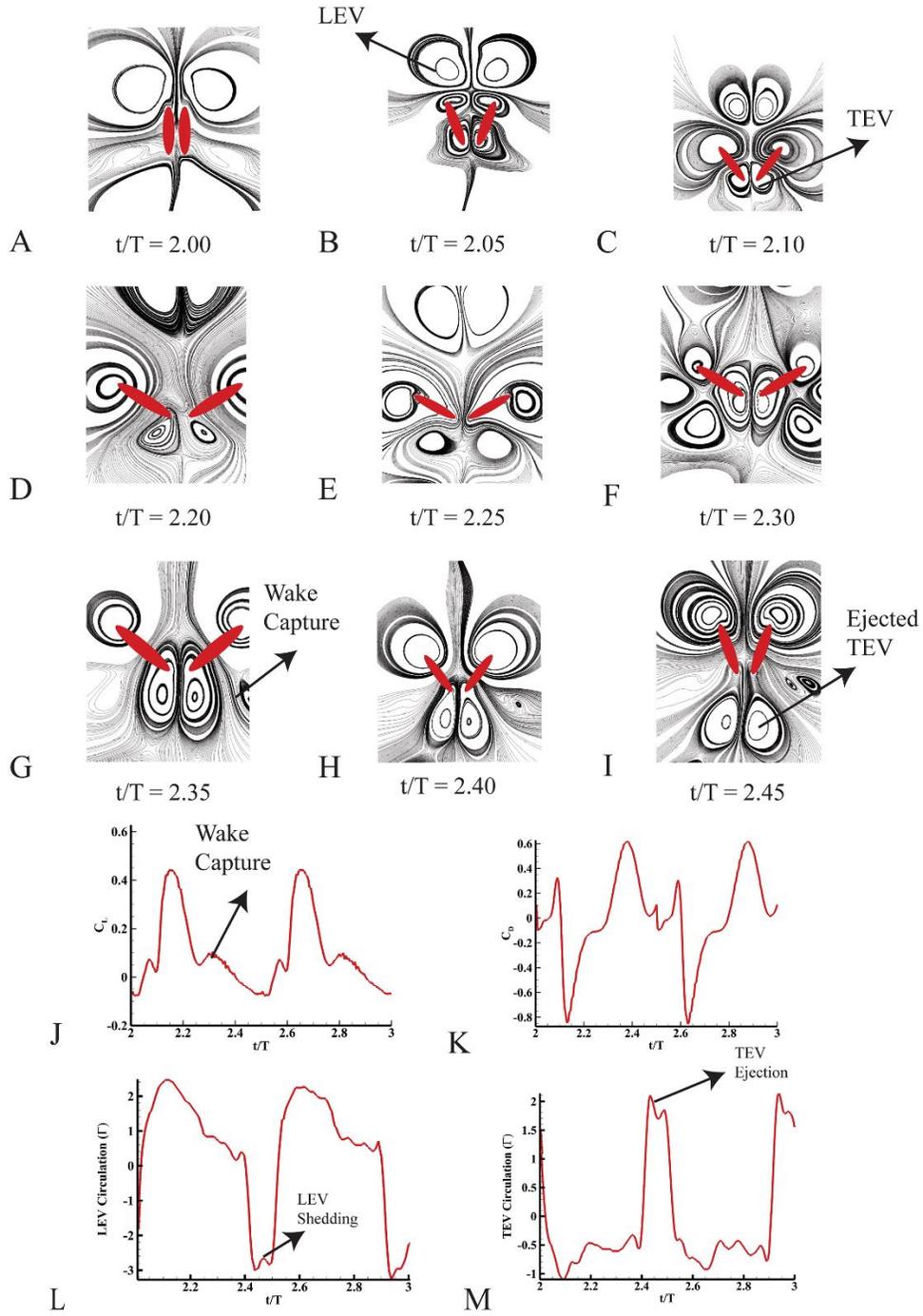

**Fig. 6:** Figure A-I corresponds to velocity streamline plot of clap and fling motion for Re =75, f=0.05Hz, J=1.67E-3. J-K represents corresponding time histories of the lift and drag coefficient. L-M shows the circulation history of both LEVs and TEVs. Here Circulation ($\Gamma$) is a non-dimensionalised entity.

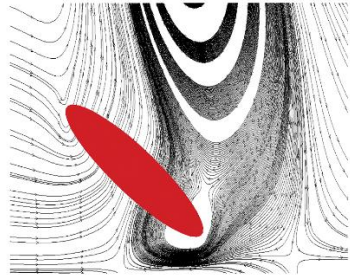

**L = 2.000**  **Γ=0.5478**

**a) Re = 75, f = 0.05 Hz, J = 2**

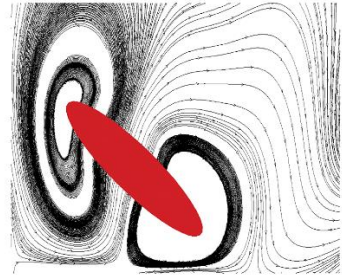

**L = 9.0553**  **Γ=1.639**

**b) Re = 75, f = 0.05 Hz, J = 1.67E-3**

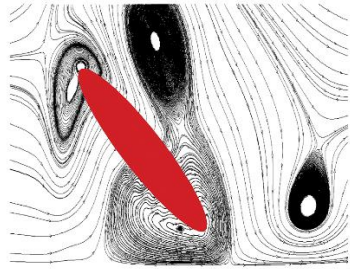

**L = 12.5499**  **Γ=1.001**

**c) Re = 75, f = 0.25 Hz, J = 2**

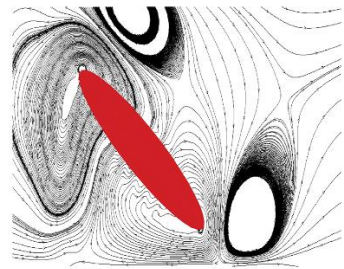

**L = 11.6190**  **Γ=0.9561**

**d) Re = 75, f = 0.25 Hz, J = 1.67E-3**

**Fig.7:** Plot comparing the size of LEV at different frequency and advance ratio for Re 75

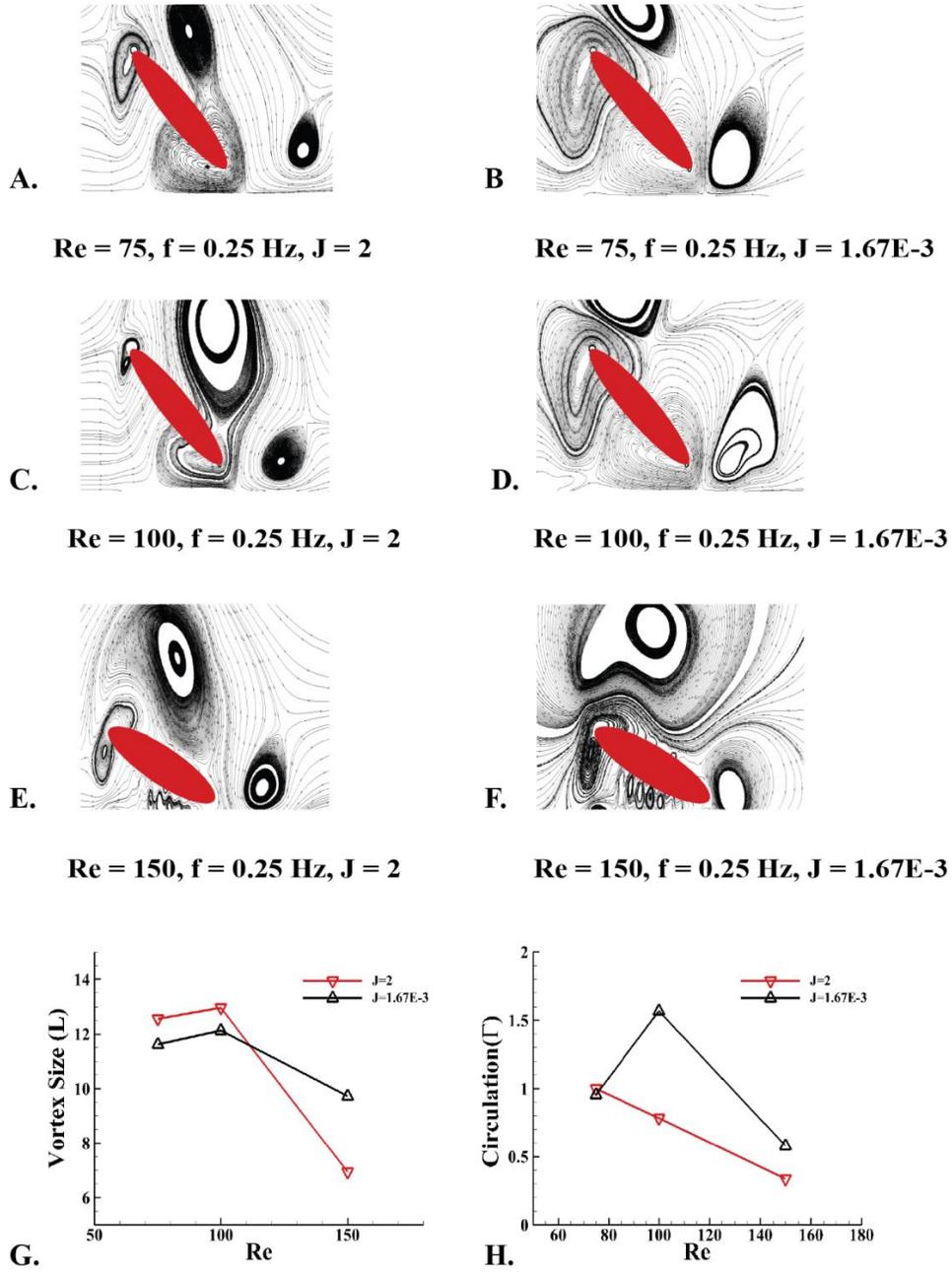

**Fig.8:** Plot A-F compares the size of LEV at different Re and advance ratio (at frequency f = 0.25Hz). G-H compares the vortex size and strength versus Reynolds number.

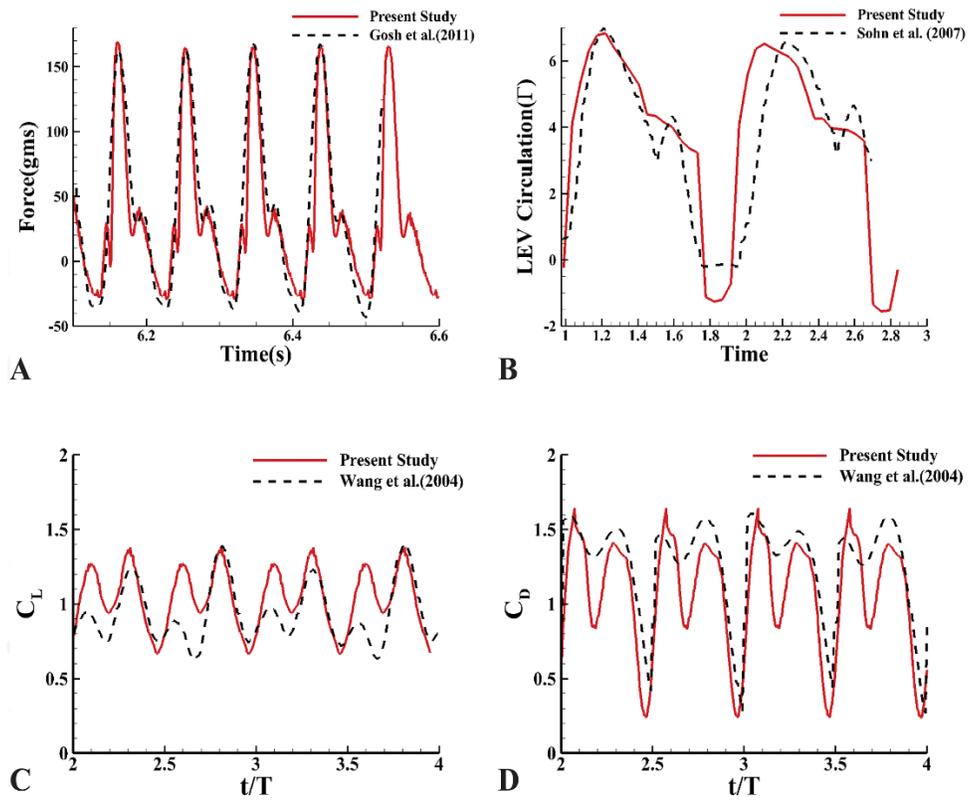

**Fig 9:** Comparison between experimental and numerical Simulation. Plot A compares the time history of lift force obtained from the present study with the experimental results of Gosh et al. [24]. In plot B, the time history of circulation is compared with Sohn et al. [25] computational results. Plot C-D is time history of lift and drag compared with the predictions of Wang et al. [23].

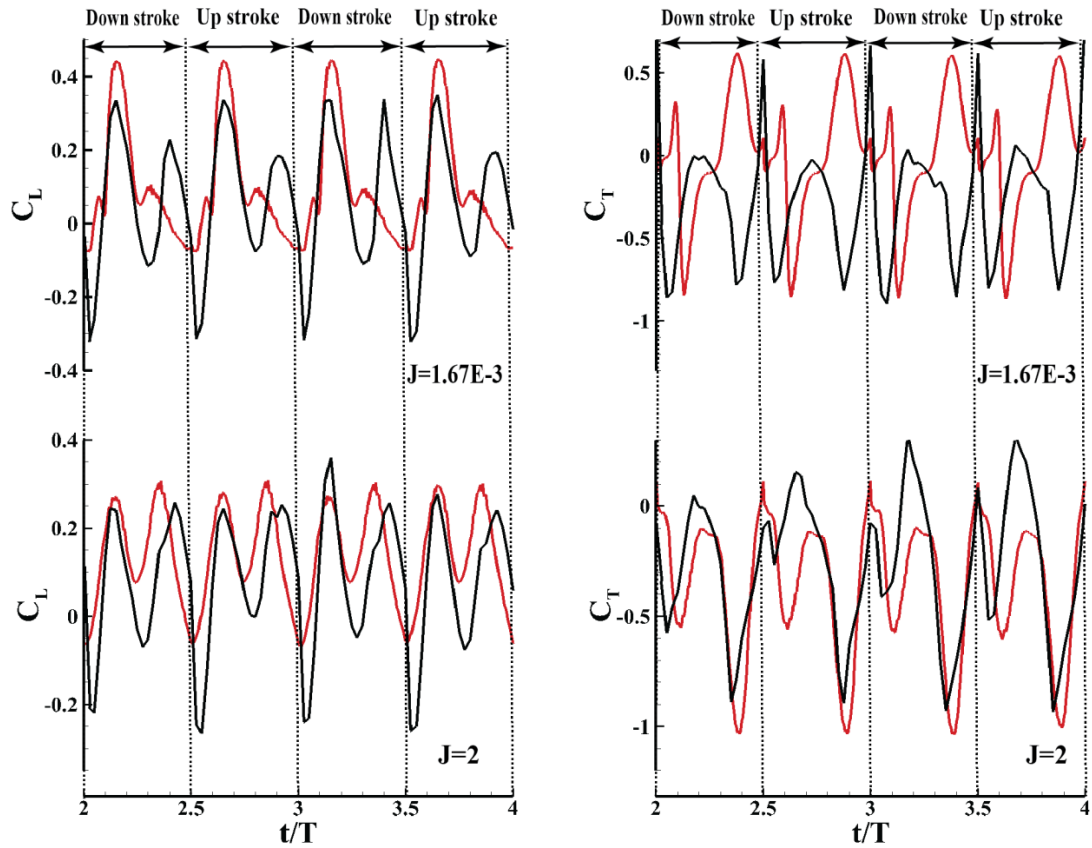

**Fig.10:** Lift and drag coefficient variation for different flow parameters. (Red) f = 0.05Hz, (Black) f = 0.25Hz, Re=100.

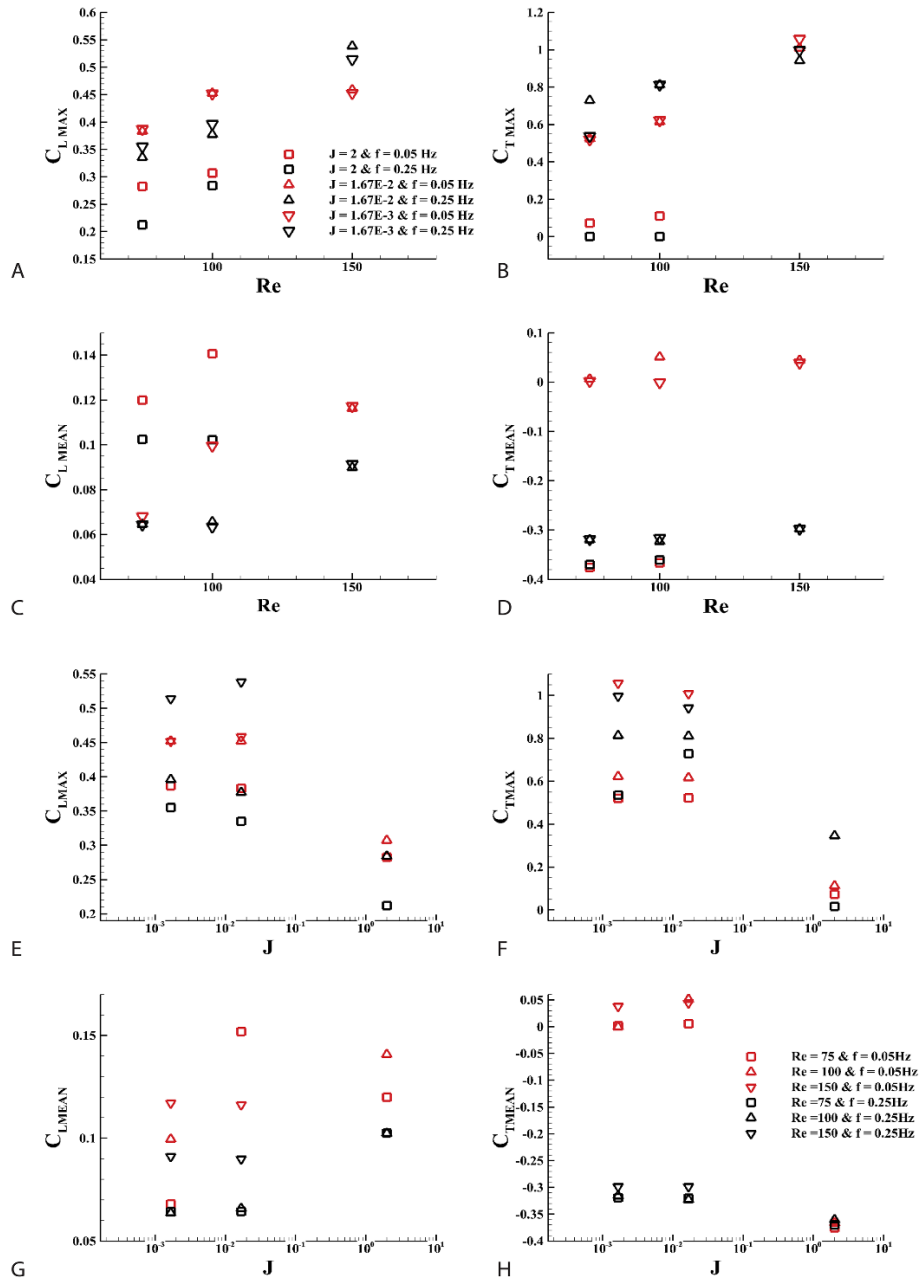

**Fig.11:** A-D shows the maximum lift & thrust coefficients and the mean lift & thrust coefficients are plotted against the Re for flow parameters involving different advance ratio and frequency. E-H shows the maximum lift & thrust coefficients and the mean lift & thrust coefficients are plotted against the advance ratio for flow parameters involving different Re and frequency.